\documentclass[12pt,aps]{revtex4}%
\usepackage{amssymb}
\usepackage{amsmath}
\usepackage{amsfonts}
\usepackage{graphicx}%
\providecommand{\U}[1]{\protect\rule{.1in}{.1in}}
\providecommand{\U}[1]{\protect\rule{.1in}{.1in}}
\providecommand{\U}[1]{\protect\rule{.1in}{.1in}}
\providecommand{\U}[1]{\protect\rule{.1in}{.1in}}
\providecommand{\U}[1]{\protect\rule{.1in}{.1in}}
\providecommand{\U}[1]{\protect\rule{.1in}{.1in}}
\begin{document}
\title{Counter-rotating Kerr Manifolds Separated by a Fluid Shell}
\author{J.P. Krisch and E.N. Glass}
\affiliation{Department of Physics, University of Michigan, Ann Arbor, MI 48109}
\date{1 July 2009}

\begin{abstract}
We describe a spheroidal fluid shell between two Kerr vacuum regions which
have opposite rotation parameters.\ The shell has a stiff equation of state
and a heat flow vector related to the rotational Killing current.\ The shell
description is useful in exploring the significance of counter-rotation in
Kerr metric matches.

\end{abstract}
\maketitle

\section{Introduction}

The Kerr vacuum metric is believed to describe the spacetime exterior to
compact rotating objects such as neutron stars \cite{Bon92}. Kerr and vacuum
Schwarzschild are considered the two most fundamental explicit solutions of
general relativity.\ While there are a variety of interior fluid models that
provide the source of the Schwarzschild mass parameter, a single interior
fluid region matched to the Kerr vacuum has not yet been discovered.\ Some of
the early investigations looked at models expressed as an expansion of the
Kerr angular momentum parameter $^{\prime}a^{\prime}.$\ Brill and Cohen
\cite{BC66} showed that to $O(a)$, a rotating spherical shell bounding a flat
interior could be matched to Kerr. De la Cruz and Israel \cite{DCI68} extended
the model to order $O(a^{2})$.\ Hartle and Thorne \cite{HT68} studied slowly
rotating relativistic stars and found that, with their general metric, any
'slowly rotating configuration' could match vacuum Kerr up to second order in
rotational speed, a result that Cohen \cite{Coh67} noticed for $O(a)$ perfect
fluids.\ McCrea \cite{McC73} and Florides \cite{Flo73, Flo75}, using a
smallness parameter $^{\prime}k^{\prime}$, discussed spherical and spheroidal
matches to $O(k)$ and $O(k^{5})$ respectively.

One of the technical problems with matching the Kerr metric to an interior is
duplicating the dependence on cos$^{2}\theta$,\ found in all of the Kerr
metric functions. The strategy of keeping the angular behavior and replacing
some or all of the functions of the radial coordinate with unknown functions
has been discussed by Krasinski \cite{Kra78}. Attempts to simplify the general
match with a perfect fluid interior have been unsuccessful \cite{Kra78,McM91}.\ 

Because some interesting astrophysical objects have spherical topology, there
are many treatments matching Kerr to matter distributions with spheroidal
boundaries. The most common boundary is $r=R_{0}$, which in the Kerr geometry
is spheroidal. Other choices are Hogan's \cite{Hog76} interior solution with
boundary $r^{2}-br+a^{2}\cos^{2}\theta=0,$ and the Gurses-Gursey \cite{GG75}
spheroid, $\ r_{0}^{2}(x^{2}+y^{2}+z^{2})-a^{2}z^{2}=r_{0}^{2}(r_{0}^{2}%
+a^{2}).$ Both models develop unphysical stress-energy in the interior
\cite{Col78}.

For models with spherical topology, one has to make a choice of a metric
interior, and often a choice of the stress-energy form.\ Disc models avoid
these choices by treating the disc as a boundary between two Kerr manifolds,
and then use the Israel \cite{Isr66} method to create a disc with
stress-energy determined by the extrinsic curvature jump across the disc.
Discs are useful models for flat galaxy-like objects with diffuse edges, and
the disc structure can be motivated by the shape of the Kerr ring
singularity.\ Israel's \cite{Isr70} source for Kerr was a layer of mass in an
equatorial disc.\ The disc had negative density and spun superluminally. Lopez
\cite{Lop81} studied rigidly rotating discs; a negative mass disc with a line
of infinite positive masses along the edges was described by Keres
\cite{Ker67}. Hamity's \cite{Ham76} disc model has no density and positive
stress.\ Hamity and Lamberti \cite{HL87} found that these disc models violated
the strong and weak energy conditions, as reflected in their mass
description.\ McManus \cite{McM91} combined a toroidal shell with a disc using
both positive and negative mass and angular momentum. Following a treatment of
counter-rotating sources for static vacuum solutions \cite{BLBK93}, Bi\u
{c}\'{a}k and Ledvinka \cite{BL93} described a disk source made of two
counter-rotating particle currents; these counter-rotating current models
appear to solve many of the matter problems occurring in the papers reviewed
by Hamity and Lamberti \cite{HL87}.\ Counter-rotation was also considered in
an early disc model by Morgan and Morgan \cite{MM69}.\ One of their examples,
using a gravitational potential communicated by Bardeen, was a disc of
uniformly counter-rotating dust.\ While they made no connection to Kerr, the
appearance of counter-rotation in a model with physical matter is interesting.
Counter-rotation has also appeared in the models of Marek and Haggag
\cite{HM81,Hag90} and Ramadan \cite{Ram04} , who described Kerr matches with a
rotating thick matter shell around a static matter core with counter-rotation
occurring in the thick shell.\ 

A component of the disc description which has produced interesting results is
counter-rotation and, for disc matches, vacuum Kerr on both sides of the
matching mass distribution. Using Kerr on both sides of a matching disk
involves a change in the sense of z\ coordinate increase (in cylindrical
coordinates).\ For example, the disk match of Bi\u{c}\'{a}k and Ledvinka
\cite{BL93} has outward directed normals on either side of the disk. The flip
in the direction of a positive normal crossing the disc, produces a non-zero
jump in the extrinsic curvature for the same metric.\ For spheroidal shells
outside of the event horizons, the interior and exterior normals are both
outward pointing, but an analogous choice is opposite senses for the azimuthal
angle.\ For a spheroidal shell with Kerr on both sides, this choice is
metrically equivalent to choosing a positive angular momentum parameter in the
exterior region and a negative parameter in the interior.\ Developing a
spheroidal model containing some of the best elements of the disc models,
counter-rotation and Kerr on both sides of the boundary, will allow the
importance of these features to be investigated for other Kerr matches.\ 

In this paper we describe a spheroidal shell between two vacuum Kerr metrics
where the two metrics differ only in the sign of parameter $^{\prime}%
a^{\prime}$. The shell is located outside of the event horizon. In the next
section we describe the metric of the shell and the observers who will
interpret its stress-energy. Using the Israel \cite{Isr66} formalism, the
shell stress-energy is developed in Section III. We find that the shell has a
stiff equation of state (not imposed), with a heat flow component following
from the Israel stress condition, and which is necessary to provide a positive
angular momentum parameter at infinity. The angular speed of the shell is
discussed in the fourth part of the paper and a simple model is also
discussed. The shell is potentially useful as a tool to explore the effects of
counter-rotation in Kerr matches. It also provides a simple model with which
to describe some general relativistic effects in astrophysical objects with
shell structure \cite{SHR08,TBC99,GOB00,CBJ+07}.

\section{Metric and Observers}

The Boyer-Lindquist (BL) form of vacuum Kerr for the interior (-) and exterior
(+) is\ \
\begin{subequations}
\begin{align}
ds_{\pm}^{2} &  =-fdt^{2}-2k_{\pm}dtd\varphi+(\Sigma/\Delta)dr^{2}+\Sigma
d\vartheta^{2}+ld\varphi^{2}\\
\Sigma &  =r^{2}+a^{2}\cos^{2}\vartheta,\text{ \ }\Delta=r^{2}+a^{2}-2mr,\\
f &  =1-2mr/\Sigma,\\
k_{\pm} &  =\pm2mar\Sigma^{-1}\sin^{2}\vartheta,\\
l &  =[(r^{2}+a^{2})^{2}-a^{2}\Delta\sin^{2}\vartheta]\Sigma^{-1}\sin
^{2}\vartheta,\\
D^{2} &  =fl+k_{\pm}^{2}=\Delta\sin^{2}\vartheta.
\end{align}
The observers on either side of a matching Israel layer \cite{Isr66} should
agree on the induced layer metric. For two Kerr vacua, the observers who agree
on\ a metric form are the zero angular momentum observers (ZAMOs), moving with
four velocity
\end{subequations}
\begin{equation}
U_{\pm}^{i}=\sqrt{l/D^{2}}[1,0,0,\Omega_{\pm}],\text{ \ \ }\Omega_{\pm}%
=k_{\pm}/l.
\end{equation}
The metric for the ZAMOs is%
\begin{equation}
dh^{2}=-(D^{2}/l)dt^{2}+(\Sigma/\Delta)dr^{2}+\Sigma d\vartheta^{2}%
+ld\psi_{\pm}^{2}\label{zamo-met}%
\end{equation}
where their angular coordinate $\psi_{\pm}$ is related to the BL $\varphi$ by
\begin{equation}
d\psi_{\pm}=d\varphi-\Omega_{\pm}dt.\label{psi-to-phi}%
\end{equation}
The boundary surface between the two metrics is the spheroidal surface defined
by $r=R_{0}$ with normal vector
\begin{equation}
n_{i}=[0,\sqrt{\Sigma/\Delta},0,0].
\end{equation}
The two ZAMO observers agree on the shell metric Eq.(\ref{zamo-met}), with the
jump in the extrinsic curvature providing the shell stress-energy \cite{Isr66}.\ 

\section{Stress-Energy}

The Israel stress-energy \cite{Isr66}, $S_{ij}$, is formed from jumps in the
extrinsic curvature, $K_{ij}$, with
\begin{equation}
-8\pi S_{ij}=\ <K_{ij}>\ -\frac{1}{2}h_{ij}<K_{k}^{k}>
\end{equation}
and extrinsic curvature
\[
K_{ij}=n_{a;b}e_{i}^{a}e_{j}^{b}.
\]
$e_{i}^{a}$ are the tangents to the layer. The extrinsic curvature components
are, with $\digamma=\sqrt{\Delta}/(2\sqrt{\Sigma}),$
\begin{subequations}
\begin{align}
K_{tt}^{\pm} &  =-\digamma\lbrack\partial_{r}f+2\Omega_{\pm}\partial_{r}%
k_{\pm}-\Omega_{\pm}^{2}\partial_{r}l]\\
K_{\vartheta\vartheta}^{\pm} &  =\digamma\partial_{r}\Sigma\\
K_{\psi\psi}^{\pm} &  =\digamma\partial_{r}l\\
K_{t\psi}^{\pm} &  =-\digamma\lbrack\partial_{r}k_{\pm}-\Omega_{\pm}%
\partial_{r}l]=-l\digamma\partial_{r}(k_{\pm}/l)
\end{align}
Only the component $K_{t\psi}$, which is odd in $`a`$, will contribute to the
jump.
\end{subequations}
\begin{align}
&  <K_{i}^{i}>\ =0\label{trace-K}\\
&  <K_{t\psi}>\ =l\digamma(\partial_{r}\Omega_{-}-\partial_{r}\Omega_{+})
\end{align}
with $\Omega_{-}=-\Omega_{+}$%
\begin{equation}
<K_{t\psi}>\ =-2l\digamma\partial_{r}\Omega_{+}%
\end{equation}
The associated stress-energy is%
\begin{equation}
8\pi S_{t\psi}=-<K_{t\psi}>\ =2l\digamma\partial_{r}\Omega_{+}\ .
\end{equation}
This stress-energy can be analyzed using an anisotropic fluid with heat flow,
$q_{i}=[q_{t},0,0,q_{\psi}].$
\begin{equation}
8\pi S_{ij}=\sigma V_{i}V_{j}+P_{(\psi)}\Psi_{i}\Psi_{j}+q_{i}V_{j}+q_{j}V_{i}%
\end{equation}
where $\Psi^{i}$ is the unit vector for the azimuthal coordinate$\ $and
$V^{i}$ the unit velocity of the layer.%
\begin{align}
V_{\pm}^{i} &  =l^{-1/2}[(D/l)^{2}-(\omega_{\pm}-\Omega_{\pm})^{2}%
]^{-1/2}\ [1,0,0,\omega_{\pm}-\Omega_{\pm}]\\
\Psi_{\pm}^{i} &  =(l^{1/2}/D)[(D/l)^{2}-(\omega_{\pm}-\Omega_{\pm}%
)^{2}]^{-1/2}\ [(\omega_{\pm}-\Omega_{\pm}),0,0,(D/l)^{2}]
\end{align}
with $\omega_{\pm}=d\varphi/dt$ the exterior/interior rotational speed for the
layer with BL angle $\varphi$.$\ $Different $\omega_{\pm}$ on either side of
the layer allows models that will include counter-rotation in the layer
motion.\ The density, pressures, and heat flow arising from the stress-energy
are%
\begin{align}
\sigma_{\pm} &  =P_{(\psi)\pm}=\frac{2(\omega_{\pm}-\Omega_{\pm})}%
{(D/l)^{2}-(\omega_{\pm}-\Omega_{\pm})^{2}}\sqrt{\frac{\Delta}{\Sigma}%
}\ \partial_{r}\Omega_{+}\\
P_{(\vartheta)} &  =0\\
q_{t\pm} &  =-(\omega_{\pm}-\Omega_{\pm})q_{\psi_{\pm}}\\
q_{i\pm}\Psi_{\pm}^{i} &  =-\sqrt{\frac{\Delta}{\Sigma}}\left[  \frac
{l\partial_{r}\Omega_{+}}{D}\right]  \frac{(D/l)^{2}+(\omega_{\pm}-\Omega
_{\pm})^{2}}{(D/l)^{2}-(\omega_{\pm}-\Omega_{\pm})^{2}}%
\end{align}
All stress-energy parameters are evaluated at a shell radius exterior to
the\ outer event horizon, $R_{0}>m+\sqrt{a^{2}-m^{2}}.$ \ Note that the stiff
equation of state is not imposed but arises from the stress-energy content.

\section{Rotational Speed}

The derivative of the exterior ZAMO angular speed is negative: $\partial
_{r}\Omega_{+}<0$ for $R_{0}$ outside of the first event horizon at $\Delta
=0$. In the exterior $\omega_{+}<k_{+}/l$ is necessary for positive
density.\ This is a smaller value than the maximum allowed rotational speed
for an exterior Kerr observer \cite{MTW70}, $(\omega_{+})_{\text{max}}%
=k_{+}/l+D/l$.\ A stricter condition is set on $\omega_{-}$. For positive
density, $\omega_{-}$ must be negative and with size larger than $\left\vert
k_{+}/l\right\vert .$\ The Kerr constraint bounds $\omega_{-}$ such that
$(\omega_{-})_{\text{min}}=-(k_{+}/l+D/l).$\ The conditions on $\omega_{\pm}$
for positive density with the Kerr constraints are%
\begin{align}
k_{+}/l-D/l\leq\omega_{+}<k_{+}/l &  \\
\omega_{-}<0 &  \nonumber\\
k_{+}/l<\left\vert \omega_{-}\right\vert \leq k_{+}/l+D/l &  \nonumber
\end{align}
$\omega_{+}$ can be positive or negative, with counter-rotation developing if
it is positive. One can find a relation between $\omega_{+}$ and $\omega_{-}$
by considering density relations. Equating the densities, $\sigma_{+}%
=\sigma_{-}$, we have \
\begin{equation}
\frac{(\omega_{+}-\Omega_{+})}{D^{2}-l^{2}(\omega_{+}-\Omega_{+})^{2}}%
=\frac{(\omega_{-}-\Omega_{-})}{D^{2}-l^{2}(\omega_{-}-\Omega_{-})^{2}}.
\end{equation}
Defining a scaled angular speed $\bar{\omega}_{\pm}=\omega_{\pm}/\Omega_{+},$
this can be written as%
\begin{equation}
D^{2}(\bar{\omega}_{+}-\bar{\omega}_{-}-2)=k_{+}^{2}(\bar{\omega}_{+}%
-1)(\bar{\omega}_{-}+1)(\bar{\omega}_{-}-\bar{\omega}_{+}+2).
\end{equation}
The two possible conditions on $\bar{\omega}_{\pm}$ are%
\begin{align}
\bar{\omega}_{+} &  =\bar{\omega}_{-}+2,\\
(1-\bar{\omega}_{+})(1+\bar{\omega}_{-}) &  =D^{2}/k_{+}^{2}\ .\nonumber
\end{align}
The first condition establishes a relation between the two rotational speeds.
It is not possible to satisfy the second condition for positive density
layers.\ In the region%
\begin{equation}
0<\bar{\omega}_{+}<1,\text{ \ }1<\left\vert \bar{\omega}_{-}\right\vert <2
\end{equation}
the layer is observed to rotate in the same direction as the corresponding
counter-rotating ZAMO. The functional form of $\omega_{\pm}$ is not
determined, but some convenient choices can be made to explore the density.
For example, the structure of the density suggests a simplifying choice: \
\begin{align}
\omega_{+}-\Omega_{+} &  =-n(D/l)\Sigma^{\gamma}\sin\theta,\label{om+}\\
\bar{\omega}_{+} &  =1-n\frac{D\Sigma^{\gamma}\sin\theta}{k}=1-n\frac
{\Sigma^{\gamma+1}\sqrt{\Delta}}{2mar}.\nonumber
\end{align}
The factor $\Sigma^{\gamma}$ is added to facilitate a discussion (in the last
part of the paper) of the shell behavior as the ring singularity is
approached. A constraint on the size of the shell is set as a function of the
model parameters:\
\begin{equation}
0<n\frac{\Sigma^{\gamma+1}\sqrt{\Delta}}{2mar}<1.
\end{equation}
The density is%
\begin{align}
\sigma_{+}=n\frac{4ma\Sigma^{\gamma-3/2}\sin^{2}\theta\lbrack3r^{4}+r^{2}%
a^{2}(2-\sin^{2}\theta)-a^{4}\cos^{2}\theta]}{[(r^{2}+a^{2})^{2}-a^{2}%
\Delta\sin^{2}\theta](1-n^{2}\Sigma^{2\gamma}\sin^{2}\theta)} &  \label{sig}\\
\text{at the equator }\theta=\pi/2\text{ and }\Sigma=r^{2} &  \nonumber\\
(\sigma_{+})_{equator}=n\frac{4mar^{2\gamma-2}(3r^{2}+a^{2})}{(r^{3}%
+a^{2}r+2ma^{2})(1-n^{2}r^{4\gamma})} &  \label{sig-eq}%
\end{align}
The equatorial density is a smoothly increasing function of $r$ coming in
toward the first event horizon ($\Delta=0).$ As a function of polar angle, the
density is peaked around the equator, decreasing to zero at the poles.\ To an
exterior observer the mass distribution is suggestive of the ring singularity.
However, with opposite rotation parameters in the interior and exterior
metrics, the shell is a source of angular momentum and mass in addition to the
ring contributions.\ Using the same form for the rotational speed, the size of
the heat flow vector and its components are%
\begin{align}
q &  =\sqrt{q_{i}q^{i}}=q_{i\pm}\Psi_{\pm}^{i}\\
q_{i\pm}\Psi_{\pm}^{i} &  =\frac{2ma\sin\theta}{\Sigma^{3/2}}\left[
\frac{3r^{4}+r^{2}a^{2}(2-\sin^{2}\theta)-a^{4}\cos^{2}\theta}{(r^{2}%
+a^{2})^{2}-a^{2}\Delta\sin^{2}\theta}\right]  \left(  \frac{1+n^{2}%
\Sigma^{2\gamma}\sin^{2}\theta}{1-n^{2}\Sigma^{2\gamma}\sin^{2}\theta}\right)
\nonumber
\end{align}%
\begin{align}
q_{t} &  =n(D/l)(\Sigma^{\gamma}\sin\theta)q_{\psi_{\pm}}\\
q_{\psi_{\pm}} &  =\frac{q_{i\pm}\Psi_{\pm}^{i}\sqrt{l}}{[1-n^{2}%
\Sigma^{2\gamma}\sin^{2}\theta]^{1/2}}\ \nonumber
\end{align}
The heat current in the shell, like the density, is peaked around the equator
as one would expect from a rotating mass distribution. The shell is a source
of angular momentum but not the singular ring usually associated with vacuum
Kerr. The shell surrounds a spinning source whose angular momentum parameter
is opposite to the one observed at infinity.\ The Komar angular momentum of
the shell, calculated in the Appendix, is $2ma.$ The shell current angular
momentum, combined with the counter-rotating ring angular momentum, provides
the $+a$ observed at infinity. \ 

\section{Discussion}

There are parallels between the Kerr rotating disc \cite{BL93} and the Kerr
rotating shell.\ The disc is described with two sets of observers. The
$\varphi-isotropic$ observers (FIOs) see a diagonal stress-energy tensor and
interpret the matter content as two streams of particles counter-rotating with
the same speeds. These observers are rotating with respect to the disc\ ZAMOs
who see the streams counter-rotate with different speeds. The equation of
state of the disc depends on the rotational speed seen by the FIOs.

The observer sets for the Kerr shell are the ZAMOs who see a diagonal metric
with angular coordinate $\psi_{\pm},$ and the observers who use BL coordinate
$\varphi.$ The ZAMOs are counter-rotating with respect to each other. The
$\pm$ shell ZAMOs who agree on the size of the layer density set limits on
$\omega_{\pm}$. Within these limits, the BL observers can interpret the shell
as having two counter-rotating elements or not, depending on the sign of
$\omega_{+}$.\ The Kerr shell has a richer set of motions than the Kerr disc
but a very limited equation of state. The shell contains a $P=\sigma$ fluid
while the content of the Kerr disc ranges from dust to stiff matter. The Kerr
shell has heat flow, while the Kerr disc does not. \ 

The heat flow vector provides an interesting insight. If $\omega_{\pm}%
=\Omega_{\pm},$ the density, pressure, and timelike component of the heat flow
vanish. The stress-energy becomes%
\[
8\pi S_{ij}=q_{i}N_{j}+N_{i}q_{j}%
\]
with $V^{i}\rightarrow N^{i}=[l^{1/2}/D,0,0,0]$ time-like. A spatial heat
component remains%
\[
q_{\psi+}=-\frac{l^{3/2}}{D}\sqrt{\frac{\Delta}{\Sigma}}\ \partial_{r}%
\Omega_{+}\ .
\]
This heat component can be identified with the rotational Killing current
\cite{Car04}. Rewriting it in terms of the $\psi\ $Killing vector $\xi
_{(\psi)}^{i}$, we have%
\[
q_{\psi+}=-8\pi S_{ij}N^{i}\xi_{(\psi)}^{j}%
\]
which is the integrand of a Komar angular momentum calculation (see Appendix
A). The Komar mass and angular momentum of the disc and the shell are
conceptually very different.\ The disc is the source of the metric on either
side, and the Kerr mass and angular momentum parameters are usually referred
to the disc.\ The shell surrounds a spinning source whose angular momentum
parameter is opposite to that observed at infinity.\ The Komar mass of the
shell is zero and the angular momentum is $2ma$.\ Other examples with zero
Komar mass and non-zero Komar angular momentum have been discussed by Glass
and Krisch \cite{GK04}. These Komar results motivate the same Schwarzschild
mass parameter in the interior and exterior as well as conservation of angular
momentum over the whole region and reflect the result of Ansorg and Petroff
\cite{AP06,AP07} for Komar mass; that Komar quantities seem more closely
related to an object in a gravitational environment than to properties of the
source by itself.

While physical observers lie outside the first Kerr horizon, it is of interest
to ask about the shell behavior as $r$ approaches the ring singularity at
$R_{0}=0$. Balasin and Nachbagauer \cite{BN94} have developed a distributional
method for exhibiting the ring structure of the Kerr source. Using outgoing
null coordinates $[\chi,v]$ with the definition $d\psi=d\chi+(k/l)dv$ allows a
view inward toward the ring singularity, where the distributional support for
the ring has been established. The functional form for the density and heat
flow are the same as in Eq.(\ref{sig-eq}).\ The renormalization factor
$\Sigma^{\gamma}$ with $\gamma=1,$ allows a non-zero density similar to the
toroidal density discussed by McManus \cite{McM91}. The equatorial heat
current is singular. In this limit, the heat flow current overlies the ring
singularity with net angular momentum parameter $+a$, and the manifold is a
single Kerr spacetime.

The shell model developed in this paper, along with the Kerr disc models,
provide a model set that can be used to study some of the challenges of a
general Kerr match.\ It incorporates some features which have been useful in
disc models, such as counter-rotation and the same metric structure on each
side of the shell. Differences such as a stiff equation of state and heat flow
are suggestive of directions for further study in more detailed models.\ The
model could also be useful in describing some general relativistic aspects of
sources with a shell structure \cite{SHR08,TBC99,GOB00,CBJ+07}.\ 

\appendix{}

\section{Komar mass and angular momentum}

The shell that separates the two vacuum Kerr regions has a physical density,
stress, and a stiff fluid equation of state.\ The Komar mass and angular
momentum can be calculated for the layer as seen by BL observers where the
integrals are over a $t=constant$ hypersurface $\Pi$ with $N^{a}$ the normal
to $\Pi$.

Asymptotically, as the spheroidal boundary is taken to infinity, the normal
is
\[
\lim_{R_{0}\rightarrow\infty}N^{a}=\delta_{(t)}^{a}%
\]
Using the coordinate transformation, Eq.(\ref{psi-to-phi}), to evaluate the BL
stress-energy yields%
\begin{align*}
8\pi S_{(t)(t)}^{\text{BL}}  &  =-2\Omega_{+}S_{(t)(\psi)}^{\text{BL}}%
=-2k_{+}\sqrt{\frac{\Delta}{\Sigma}}\ \partial_{r}\Omega_{+}\\
8\pi S_{(t)(\varphi)}^{\text{BL}}  &  =S_{(t)(\psi)}^{\text{BL}}=\sqrt
{\frac{\Delta}{\Sigma}}\ l\partial_{r}\Omega_{+}\\
\text{trace }S^{\text{BL}}  &  =0
\end{align*}
Asymptotically we have
\begin{align*}
&  \partial_{r}\Omega_{+}\sim-\frac{6ma}{r^{4}},\\
&  8\pi S_{(t)(t)}^{\text{BL}}\sim\delta(r-R_{0})\frac{4ma\sin^{2}\vartheta
}{r}\frac{6ma}{r^{4}},\\
&  \sqrt{-g}d^{3}x\sim r^{2}\sin\vartheta drd\vartheta d\varphi.
\end{align*}
The Komar mass is%
\begin{align}
M_{k-layer}  &  =2\lim_{R_{0}\rightarrow\infty}%
{\displaystyle\int\limits_{\Pi}}
S_{ab}^{\text{BL}}N^{a}\delta_{(t)}^{b}\sqrt{-g}d^{3}x,\\
&  =2\lim_{R_{0}\rightarrow\infty}%
{\displaystyle\int\limits_{\Pi}}
S_{(t)(t)}^{\text{BL}}\sqrt{-g}d^{3}x,\nonumber\\
&  =2\lim_{R_{0}\rightarrow\infty}2\pi%
{\displaystyle\int\limits_{\Pi}}
\delta(r-R_{0})\frac{4ma\sin^{2}\vartheta}{8\pi r}\frac{6ma}{r^{4}}r^{2}%
\sin\vartheta drd\vartheta,\nonumber\\
&  =2\lim_{R_{0}\rightarrow\infty}%
{\displaystyle\int\limits_{\Pi}}
\frac{ma\sin^{3}\vartheta}{R_{0}}\frac{6ma}{R_{0}^{4}}R_{0}^{2}d\vartheta
,\nonumber\\
&  =0.\nonumber
\end{align}
The Komar angular momentum is%
\begin{align}
L  &  =-\lim_{R_{0}\rightarrow\infty}%
{\displaystyle\int\limits_{\Pi}}
S_{ab}N^{a}\delta_{(\varphi)}^{b}\sqrt{-g}d^{3}x,\\
&  =-\lim_{R_{0}\rightarrow\infty}%
{\displaystyle\int\limits_{\Pi}}
S_{(t)(\phi)}\sqrt{-g}d^{3}x,\nonumber\\
&  =-\lim_{R_{0}\rightarrow\infty}%
{\displaystyle\int\limits_{\Pi}}
-\delta(r-R_{0})\frac{6mar^{2}\sin^{2}\vartheta}{8\pi r^{4}}2\pi r^{2}%
\sin\vartheta drd\vartheta,\nonumber\\
&  =\lim_{R_{0}\rightarrow\infty}%
{\displaystyle\int\limits_{\Pi}}
\frac{6maR_{0}^{2}\sin^{2}\vartheta}{4R_{0}^{4}}R_{0}^{2}\sin\vartheta
d\vartheta,\nonumber\\
&  =2ma.\nonumber
\end{align}


\begin{thebibliography}{99}                                                                                               %


\bibitem {Bon92}W.B. Bonnor, \textit{Physical Interpretations of Vacuum
Solutions of Einstein's Equations}, Gen. Rel. Gravit. \textbf{24}, 551 (1992).

\bibitem {BC66}D.R. Brill and J.M. Cohen, \textit{Rotating masses and their
effect on inertial frames}, Phys. Rev. \textbf{143}, 1011 (1966).

\bibitem {DCI68}V. DeLaCruz and W. Israel, \textit{Spinning shell as a source
of the Kerr metric}, Phys. Rev. \textbf{170}, 1187 (1968).

\bibitem {HT68}J.B. Hartle and K.S. Thorne, \textit{Slowly Rotating
Relativistic Stars, II: Models for Neutron Stars and Supermassive Stars,
}Astrophys. J. \textbf{153}, 807 (1968).

\bibitem {Coh67}J.M. Cohen, \textit{Note on the Kerr metric and Rotating
Masses}, J. Math. Phys. \textbf{8}, 1477 (1967).

\bibitem {McC73}J.D. McCrea, \textit{Gravitational field of a uniformly
rotating sphere in third approximation}, Proc. Roy. Irish. Acad. \textbf{A73},
25 (1973).

\bibitem {Flo73}P.S. Florides, \textit{A Rotating Sphere as a Possible Source
of the Kerr Metric}, Nuov. Cim. \textbf{B13}, 1 (1973).

\bibitem {Flo75}P.S. Florides, \textit{A Rotating Spheroid as a Possible
Source of the Kerr Metric}, Nuov. Cim. \textbf{B25}, 251 (1975).

\bibitem {Kra78}A. Krasinski, \textit{Ellipsoidal space-times, sources for the
Kerr metric},\emph{\ }Ann. Phys. (N.Y.) \textbf{112}, 22 (1978).

\bibitem {McM91}D. McManus, \textit{A toroidal source for the Kerr black hole
geometry}, Class. Quan. Grav. \textbf{8}, 863 (1991).

\bibitem {Hog76}P.A. Hogan, \textit{An Interior Kerr Solution}, Lett. Nuov.
Cim. \textbf{16}, 33\ (1976).

\bibitem {GG75}M. Gurses and F. Gursey, \textit{Lorentz covariant treatment of
the Kerr-Schild geometry}.\ J. Math. Phys. \textbf{16,} 2385 (1975).

\bibitem {Col78}P. Collas, \textit{Simple Tests for Proposed Interior Kerr
Metrics}, Lett. Nuov. Cim. \textbf{21}, 68 (1978).

\bibitem {Isr66}W. Israel, \textit{Singular Hypersurfaces and Thin Shells in
General Relativity}, Nuov. Cim. \textbf{44B}, 1 (1966).\ 

\bibitem {Isr70}W. Israel, \textit{Source of the Kerr Metric}, Phys. Rev. D
\textbf{2}, 641 (1970).

\bibitem {Ham76}V. Hamity, \textit{An \textquotedblleft
interior\textquotedblright\ of the Kerr metric}, Phys. Lett. A \textbf{56}, 77 (1976).\ 

\bibitem {HL87}V. Hamity and W. Lamberti, \textit{A Family of Rotating Disks
as Sources of the\ Kerr Metric}, Gen. Rel. Gravit. \textbf{19}, 917 (1987). \ 

\bibitem {Ker67}H. Keres, \textit{Physical Interpretation of Solutions to the
Einstein Equations,} Zh. Eksper. Teor. Fiz. \textbf{52,} 768 (1967).

\bibitem {Lop81}C.A. Lopez, \textit{Rigidly Rotating Disk as a Source of the
Kerr Geometry}, Nuov. Cim. B \textbf{66}, 17 (1981).

\bibitem {MM69}T. Morgan and L. Morgan, \textit{The Gravitational Field of a
Disc}, Phys. Rev. \textbf{183}, 1097 (1969).

\bibitem {BLBK93}J. Bi\u{c}\'{a}k, D. Lynden-Bell, and J. Katz,
\textit{Relativistic disks as sources of static vacuum spacetimes}, Phys. Rev.
D \textbf{47}, 4334 (1993).

\bibitem {BL93}J. Bi\u{c}\'{a}k and T. Ledvinka, \textit{Relativistic disks as
sources of the\ Kerr metric,} \ Phys. Rev. Lett. \textbf{71}, 1669 (1993).

\bibitem {Poi04}E. Poisson, \textit{A Relativists Toolkit}, (Cambridge
University Press, Cambridge, 2004). pp 90-95

\bibitem {HM81}S. Haggag and J. Marek, \textit{A nearly perfect fluid source
for the Kerr metric}, Nuov. Cim. B \textbf{62}, 273 (1981).

\bibitem {Hag90}S. Haggag, \textit{A fluid source for the Kerr metric}, Nuov.
Cim. B \textbf{105}, 4 (1990).

\bibitem {Ram04}M.A. Ramadan, \textit{Fluid sources for the Kerr metric},
Nuov. Cim. B \textbf{119}, 123 (2004).

\bibitem {SHR08}N. Smith, K.H. Hinkle, and N. Ryde, \textit{Red Supergiants as
Potential Type IIn Supernova Progenitors:\ Spatially Resolved 4.6 micron CO
Emission around VY CMa and Betelgeuse}, arXiv/astro-ph/0811.3037.

\bibitem {TBC99}A.J. Turnball, T.J. Bridges, and D. Carter, \textit{Imaging of
the Shell Galaxies NGC474 and NGC7600 and Implications for their Formation},
Mon. Not. Roy. Astr. Soc. \textbf{304}, 967 (1999).\newpage

\bibitem {GOB00}C.D. Gill and T.J. O'Brien, \textit{Hubble Space Telescope
imaging and ground-based spectroscopy of old nova shells-I. FH Ser,V533 Her,
BT Mon, DK Lac, V476 Cyg}, Mon. Not. Roy. Astr. Soc. \textbf{314}, 175 (2000).

\bibitem {CBJ+07}G. Canalizo, N. Bennert, B. Jungwiert, A. Stockton, F.
Schweizer, M. Lacy, C. Peng, \textit{Spectacular Shells in the Host Galaxy of
the QSO MC2 1635+119}, Astrophys. J. \textbf{669}, 201 (2007).

\bibitem {MTW70}C.W. Misner, K.S. Thorne, and J.A. Wheeler,
\textit{Gravitation} (W.H. Freeman, San Francisco, 1970) p894.

\bibitem {Car04}S. Carroll, \textit{Spacetime and Geometry} (Addison Wesley,
New York, 2004) p254.

\bibitem {GK04}E.N. Glass and J.P. Krisch, \textit{Spinning up asymptotically
flat spacetimes}, Class. Quan. Grav. \textbf{21}, 5543 (2004).\ 

\bibitem {AP06}M. Ansorg and D. Petroff, \textit{Negative Komar mass of single
objects in regular, asymptotically flat spacetimes}, Class. Quan. Grav.
\textbf{23, }L81 (2006) 

\bibitem {AP07}M. Ansorg and D. Petroff, \textit{Negative Komar masses in
regular stationary spacetimes},\ \ arXiv/gr-qc/0708.3899.

\bibitem {BN94}H. Balasin and H. Nachbagauer, \textit{Distributional
energy-momentum tensor of the Kerr-Newman spacetime family}, Class. Quan.
Grav. \textbf{11}, 1453 (1994).
\end{thebibliography}
\end{document}